\documentclass[12pt]{iopart}
\begin{document}
\newcommand{\blankline}{\vskip .3cm}
\newcommand{\f}{\begin{equation}}
\newcommand{\ff}{\end{equation}}
\newcommand{\be}{\begin{equation}}
\newcommand{\ee}{\end{equation}}
\newcommand{\bea}{\begin{eqnarray}}
\newcommand{\eea}{\end{eqnarray}}

\title[Gravity's Rainbow]{Gravity's Rainbow}

\author{Jo\~ao Magueijo\dag and Lee Smolin\ddag}

\address{\dag\ Theoretical Physics,
The Blackett Laboratory, Imperial
College,  Prince Consort Road, London SW7 2BZ, UK}
\address{\ddag\ Perimeter Institute for Theoretical Physics,
Waterloo, Canada  N2J 2W9\ \ and Department of Phyiscs,
University of Waterloo}

\begin{abstract}
Non-linear special relativity (or doubly
special relativity) is a simple framework for encoding properties
of flat quantum space-time. In this paper we show how this
formalism may be generalized to incorporate curvature (leading to
what might be called ``doubly general relativity''). We first
propose a dual to  non-linear realizations of relativity in
momentum space, and show that for such a dual the space-time
invariant is an energy-dependent metric. This leads to an
energy-dependent connection and curvature, and a simple
modification to  Einstein's equations. We then examine solutions
to these equations. We find the counterpart to the cosmological
metric, and show how cosmologies based upon our theory of gravity
may solve the ``horizon problem''. We discuss the
Schwarzschild solution, examining the conditions for which the
horizon is energy dependent. We finally find the weak field limit.
\end{abstract}



\maketitle

\section{Introduction}

It is generally believed that the geometry of spacetime is
fundamentally described by a quantum theory, in which the smooth
manifold and metric of general relativity is replaced by a quantum
mechanical description. Whatever the nature of that description,
it is believed that the Planck energy $E_{Pl} = \sqrt{\hbar c^5
/G}$ plays the role of a threshold separating the classical
description from the quantum description. When probed at energies
above $E_{Pl}$ we expect that a radically new picture of spacetime
will be needed.

Several candidates for that description are under study, including
loop quantum gravity~\cite{rovelli,carlip}, string
theory~\cite{pol,stringref}, Lorentzian dynamical 
triangulations~\cite{triang},
non-commutative geometry~\cite{noncom}, condensed matter 
analogues~\cite{condens}, etc. A key
issue that arises in such studies is the nature of the transition
between the fundamental quantum description and the effective low
energy description in terms of classical general 
relativity~\cite{amelstat,gli,leejoao}. This
is needed not only theoretically. It has recently become clear
that astronomical and cosmological observations make it possible
to probe the leading effects in $l_{Pl} = 1/E_{Pl}$ around the
classical limit (from now on we shall set
$\hbar=c=1$). These include possible modifications of the
energy-momentum relations\be\label{disp}E^2-p^2=m^2\ee For
instance to leading order they could take the form, \f E^2 = p^2 +
m^2 + \alpha l_{Pl} E^3 + ... \ff where $\alpha$ is a 
dimensionless constant of
order unity. The effects of such terms may, on a variety of
assumptions, be observed, or even---on some assumptions---ruled
out in observations including tests of thresholds for ultra high
energy cosmic rays~\cite{review,crexp,amel,amel1,liouv,cosmicray,leejoao1} 
and Tev
photons~\cite{gammaexp}, a possible energy dependence of the speed
of light observable in gamma ray bursts~\cite{amel}, as well as
tests involving synchrotron
radiation~\cite{crab,amecritic,replyac} and nuclear physics
experiments~\cite{nuke}. Related effects may also be detectable in the near
future in CMB observations~\cite{mersi}.

Key to the understanding of such effects is the fate of global Lorentz
symmetry in the classical
limit of the quantum theory of gravity. While many physicists 
expect that Lorentz invariance remains unbroken, this may be 
utopic. For one thing,
Lorentz invariance, unlike
rotational invariance, involves an unbounded parameter. No
matter how well
it has been tested, up to some
boost parameter $\gamma$, there is
always an infinite range to go. 
More seriously, Lorentz
invariance cannot
be a fundamental symmetry of the quantum theory of gravity, for it plays
no fundamental
role in classical general
relativity. From the point of view of general relativity, global Lorentz
invariance is
only an accidental symmetry
of a particular solution to the field equations. Thus, whether it is
broken or modified, global Lorentz
invariance cannot be an exact symmetry of the theory; rather, it is an
approximate symmetry that emerges at low energies from the
quantum theory of gravity.

There are then four main possibilities for how Lorentz invariance may be
realized when effects to leading order
in $l_{Pl}E$ are included,  where $E$ is the energy of some quanta
observed by an inertial observer:

\begin{enumerate}

\item{}Lorentz invariance and the relativity of inertial frames are
maintained exactly, without modification, but the fundamental 
matter degrees of freedom may be non-pointlike. 
This is {\it assumed} in the construction of perturbative string
theory (see, however~\cite{defst}) and other perturbative 
approaches to quantum gravity. Given
the observational possibilities just mentioned, this  will likely
be the first assumption involved in the construction of string theory
to be tested experimentally.

\item{}There is a preferred frame, observable in effects at leading
order in $l_{Pl}E$, which break Lorentz
invariance~\cite{amel,amel1,liouv} (for more extreme examples
see~\cite{am,vslrev}). 
This generally leads to a phenomenology in which there
are corrections to the energy-momentum relations of the form
(\ref{disp}). Other relations, such as energy-momentum
conservation, remain as before, so long as they are computed in
the preferred frame. It has been suggested that this scenario is,
on certain assumptions regarding dynamics, already ruled out
by present observations~\cite{crab,replyac,myers}.

\item{}Lorentz invariance is spontaneously broken by a vector
field taking on a vacuum expectation value, leading to a (possibly
locally varying) preferred frame. This option has been explored in
\cite{moffat93,gzkmoffat}.

\item{}The relativity of inertial frames is maintained, but the
transformation laws now act on momentum space non-linearly,
picking up new terms in $l_{Pl}E$~\cite{amelstat,gli,leejoao,leejoao1}. 
As a consequence the invariant
norm on momentum space is no longer bilinear, leading to corrections
to the usual energy-momentum relations (\ref{disp}). This proposal
is variously known as non-linear, deformed, or doubly special relativity
(DSR)~\footnote{
It has been argued (e.g.~\cite{ahl1}) that the non-linearities in 
DSR can be removed by making a redefinition of energy and momenta. 
This has been refuted by two developments: the example of 2+1 
gravity~\cite{2p1} and the realization that DSR's phase space
has an {\it invariant} curvature~\cite{desit}. Redefinitions of
energy and momentum are also unphysical for certain position
space formulations~\cite{djj}.}. It has other
non-trivial consequences, which are explored in several papers
(e.g. \cite{leejoao,leejoao1}); for example the laws of energy and
momentum conservation continue to hold in all inertial frames, but
they are non-linear~\cite{leejoao1,desitter}.
One motivation for this proposal is that the energy scale $E_{Pl}$
which denotes the boundary between the quantum and classical
description of spacetime, may become an invariant, in the sense
that all inertial observers agree on whether a particle
has more or less than this energy.  This resolves an otherwise
troubling issue: how a threshold between a quantum and
a classical description can depend on the motion of an observer.

\end{enumerate}

This paper is a contribution to the last proposal. We examine
the question of how the modification of special relativity
proposed in \cite{leejoao,leejoao1} can be extended to general
relativity. Our inquiry is meant to hold in the sense of the
leading corrections to a limit in which the classical spacetime
description is recovered for low energy quanta from a purely
quantum spacetime geometry. Thus we are concerned with the
effects on the propagation of quanta with energies smaller than
$E_{Pl}$ but whose wavelengths are much shorter than the local
radius of curvature.

Our main result is that we do find that there is a
sensible modification of the principles and equations of general
relativity that makes sense in this regime. This is characterized
by the feature that the geometry of spacetime becomes energy
dependent. Thus, quanta of different energies see different
classical geometries. These classical geometries share the same
inertial frames, and so the equivalence principle can be
maintained, in a modified form proposed below. But measurements of
distance and time now pick up a new dependence on the energy of
the quanta used in the measurements.

This conclusion is in part motivated by~\cite{djj}. 
Deformed special relativity
was initially proposed in momentum space. With loss of linearity
the dual position space no longer mimicks momentum space. One
possible reconstruction of position space~\cite{djj} leads
to spacetime positions subject to energy dependent transformation 
laws. Concomitantly,  the metric become energy dependent. 

One consequence of this new picture is that the velocity of light
and other massless quanta naturally becomes energy dependent.
Thus, when we turn to the study of cosmological models we find
naturally the variable speed of light cosmologies (VSL),
previously proposed in \cite{moffat93,am} (see~\cite{vslrev}
for a review of the field). It has been speculated
that there may be a connection between modified or doubly
special relativity and variable speed of light
cosmologies~\cite{ncvsl,ncinfl,pogo}. What we show here is that
this connection does indeed follow from a natural extension to
general relativity.

\section{Deformed special relativity and the rainbow metric}
Deformed or doubly special relativity is a class of theories that
implement a modified set of principles of special relativity.
These are
\begin{enumerate}

\item{}The relativity of inertial frames.

\item{}In the limit $ E/E_{Pl} \rightarrow 0$ the speed of a
photon or massless quanta goes to a universal constant, $c$, which is
the same for all inertial observers.

\item{}$E_{Pl}$ in the above condition is also a universal constant,
and is the same for all inertial observers.

\end{enumerate}
As a result the invariant of energy and momentum is modified to
\f\label{dispdef}
E^2 f^2 {\left(E/ E_{Pl}\right)} - p \cdot p g^2{\left(E/
E_{Pl}\right)}   = m^2
\ff
This can be realized by the action of a non-linear map from momentum
space to itself, denoted, $U: {\cal P}\rightarrow {\cal P}$ given by
\f\label{U}
   U \cdot (E , p_i) = (U_0,U_i)= {\left(f{\left(E\over E_{Pl}\right)} E 
,g{\left(E\over E_{Pl}\right)}p_i \right)}
\ff which implies that momentum space has a non-linear norm, given
by \f |p|^2 =  \eta^{ab} U_a(p ) U_b(p) \ff This norm is preserved
by a non-linear realization of the Lorentz group, given by
\f\label{gen}
   \tilde{L}_a^b = U^{-1} \cdot L_a^b \cdot U
\ff
where $ L_a^b$ are the usual generators.
Some examples of theories of this type are described in
\cite{leejoao,leejoao1}. The presence of a singularity in
$U$ marks the emergence of an invariant energy scale, as
required by principle (iii) above \cite{leejoao1}.

These theories are typically formulated in momentum space, and to
discuss how general relativity might be set up, one must first
discuss how to identify a dual space representing
positions~\cite{djj}. Since the momentum transformation laws are
no longer linear, the definition of a dual space is non-trivial. A
number of different answers have been proposed to the question of
what is the modified spacetime geometry consistent with deformed
or doubly special relativity. Among the possible answers are
non-commutative geometry, for example, $\kappa$ deformed Minkowski
spacetime~\cite{velnonc,velnonc1}.

We take the view here that this is the wrong question to ask and
that, instead, there is no single classical spacetime geometry
when effects of order $l_{Pl}E$ are taken into account. Instead,
we propose that classical spacetime is to leading order in
$l_{Pl}$ represented by a one parameter family of metrics,
parameterized by the ratio $E/E_{Pl}$. That is, just as the
properties of a material may depend on the energy of a phonon in
it, we take the view that the geometry of spacetime may depend on
the energy of a particle moving in it. Thus, spacetime geometry
has an effective description; in the language of the
renormalization group, geometry ``runs." Hence there is no single
spacetime dual to momentum space; the dual to momentum
space is the energy dependent family of metrics.

We stress that the argument $E$ in the metric $g_{ab}(E)$
 is {\it not} the energy of
the space-time. Instead it is the scale
at which the geometry of spacetime is probed. That is,
{\it if a freely falling inertial observer uses the motion of a particle, or
system of particles, to measure the geometry of the spacetime, $E$ is the
total energy of that particle or system of particles, as measured by that
observer.} The construction should be such that the metric is
co-variant with regards to the dual of the non-linear representation
of the Lorentz group (\ref{gen}) encoding deformed special relativity.
That such covariance may be achieved is proved by the second example
below.

Another way of describing these properties is by saying that in the
absence of gravity spacetime has an energy-dependent geometry in
the sense that particles of energy $E$ move in a geometry given by
an energy dependent set of orthonormal frame fields \f e_0 =
f^{-1} (E/E_{Pl}) \tilde{e}_0 , \ \ \ \ \ e_i = g^{-1} (E/E_{Pl})
\tilde{e}_i \ff where the tilde quantities refer to the energy
independent frame fields that specify the geometry probed by low
energy quanta. The metric given by
 \f\label{flatg}
g(E) = \eta^{ab} e_a \otimes e_b
 \ff
 is flat for all $E$. It can be considered to be a one parameter
 family of flat metrics, parameterized by $E$. The metrics share
 the same set of inertial frames, but due to the scalings, generally they
do
 not share all their geodesics; instead geodesics are generally
 energy dependent.  This is equivalent to saying that the energy
 momentum relations are modified, so they are no longer quadratic.
 We refer to such a one parameter energy-dependent family of
 metrics as a single "rainbow metric", and the metric above is the
 flat rainbow metric.

\subsection{Two well-known motivations: string theory and loop quantum gravity}
The statement made in the previous Section is a postulate, and 
as such it is not derived from anything. 
But, even if it is not derived from anything, there are examples in which
such scale dependent geometries are present in quantum theories of
gravity. One well studied example comes from conformal field theory and
string theory. The target space metric $G_{AB}$ in the standard string action
\be
I= \int \sqrt{h}h^{ij} G_{AB} \partial_i \Phi^A \partial_j \Phi^B
\ee
is in fact energy dependent, and has been treated as such since early
works of Friedan and Polyakov. That is, if one regards the theory as
fundamentally two dimensional, the target space metric $G^{AB}$ is just a
matrix of coupling constants, and as any coupling constants in
quantum field theory, these run, and become dependent on the
ratio of an energy scale $E$ to the cut off scale $M$.

This may seem nonsensical if one regards the geometry of spacetime as
primary. However, in the currently well studied theories of quantum
gravity, including loop quantum gravity and string theory, the classical
geometry of spacetime is not primary. Rather, it emerges as a low energy
coarse grained description of a very different quantum geometry. In loop
quantum gravity this is completely explicit. The classical metric is as
much an emergent quantity as the thermodynamic variables in ordinary
statistical physics. As such,  the spacetime
metric must depend on the ratio of a cutoff scale, $M$, to the scale probed,
$E$;  hence $g_{ab} (E/M)$. The usual classical metric is only defined as the
limit
\be
{\lim_{E \rightarrow 0}} g_{ab} (E/M) = g_{ab}^{classical}
\ee
The energy dependence of $g_{ab}$ is then not just consistent with the
present viewpoint in quantum gravity, it is required by it.

\subsection{Another motivation}

The rainbow metric is closely related to the method developed in 
\cite{djj} for constructing position space in deformed special 
relativity. In this approach one requires 
that free field theories in flat space-time have plane wave
solutions (examples of such field theories are presented in~\cite{djj}
and elsewhere), even though the 4-momentum they carry satisfies
deformed dispersion relations (\ref{dispdef}). For this to 
be possible the contraction between position and 
momentum (providing the phase for such waves) must remain linear (so 
that the waves remain ``plane''). That is:
\be dx^ap_a=dx^0p_0+
dx^ip_i \ee
If momentum transforms non-linearly (from a non-linear
action derived from the generators Eq.~(\ref{gen})) then this
requires that the $d x^a$ tranformation be energy dependent, as
explained in \cite{djj}. 
It is not difficult to show, that for a $U$ of the form (\ref{U})
the space-time dual has invariant: \f\label{ds2} ds^2 = -
{(dx^0)^2\over f^2} + {(dx^i)^2\over g^2} \ff
Thus, the dual space $d x^a$ is endowed
with an energy dependent quadratic invariant, that is, an
energy-dependent metric.

This example further elucidates the meaning of $E$ in $g_{ab}(E)$. 
If a given observer sees a particle (or plane wave, or
wave-packet) with energy $E$, then he concludes that this  
particle feels the metric $g_{ab}(E)$. If this particle's
energy is $E'\neq E$ for a different observer, then the latter will
assign to the particle a different metric, $g_{ab}(E')$. 
It may also happen that 
the first observer sees a different particle
in the same place but with a different energy - and accordingly
assign to it a different metric. So not only different observers
may see a given particle being affected by different metrics, but the
same observer may assign different metrics to different
particles moving in the same region at the same time. 
This is {\it required} by covariance, once we allow for
a non-linear representation of the Lorentz group in momentum space.

This argument, valid in flat space-time, carries over locally
to curved space-times using the equivalence principle as detailed
in the next Section.

\section{The deformed equivalence principle}
Having defined the position space dual to deformed relativity
in momentum space, we are now ready to consider general relativity.
We start by stating the deformed equivalence and
correspondence principles:

\begin{itemize}

\item{}{\bf Modified equivalence principle} 

Consider a region of
spacetime in which the radius of curvature $R$ is much larger than
$E_{Pl}^{-1}$.  Then freely falling observers, making measurements
of particles and fields with energies $E$ which satisfy $1/R << E
<< E_{Pl}$ observe the laws of physics to be, to first order in
$1/R$, the same as in modified special relativity.  Hence freely
falling observers to first order in $1/R$ can describe themselves
as being inertial observers in rainbow flat spacetime describe by
(\ref{flatg}). In particular, they use a family of energy
dependent orthonormal frames given locally by (\ref{flatg}).

\item{}{\bf Correspondence principle} 

In the limit 
$E/E_{PL} \rightarrow 0$
ordinary classical general relativity is recovered.

\end{itemize}
We insist on the restriction $1/R << E $
because otherwise we may have to
take into account
terms in $R(\partial p /p)$ coming
from the fact that the wavelength of a
quanta is not much smaller
than the radius of curvature.
The upper limit $E<< E_{Pl}$ comes from
the
expectation that the geometry
of quantum spacetime does not have a smooth, classical description for
energies of Planck scales and
higher.

The modified equivalence principle imples that spacetime is described by
a one parameter
family of metrics given in terms of a
one parameter family of orthonormal
frame fields
\f\label{metric}
  g (E) = \eta^{ab} e_a (E) \otimes e_b (E)
\ff
where the energy dependence of the frame fields is given by
\f
   e_0 (E) = { 1\over f(E/E_{Pl}) } \tilde{e}_0    , \ \ \
   e_i (E) = { 1\over g(E/E_{Pl}) } \tilde{e}_i
\ff
The correspondence principle then requires that
\f
   {\lim_{E/E_{Pl} \rightarrow 0}} f ( E/E_{Pl}) \rightarrow 1, \ \
\ff
and likewise for $g(E/E_{Pl})$.

This then leads to a one parameter family of connections $\nabla
(E)_\mu$ and curvature tensors $R(E)_{\mu \nu \lambda}^\sigma$
defined by the usual formulas. One defines also a one parameter
family of energy-momentum tensors $T_{\mu \nu}(E)$ and  the
Einstein equations are replaced by a one parameter family of
equations \f
   G_{\mu \nu} (E) = 8 \pi G(E) T_{\mu \nu}(E) + g_{\mu \nu}\Lambda (E)
\ff
where $G(E)$ is an energy dependent Newton's constant, defined so that
$G(0)$ is the physical Newton's constant. The energy dependence of
$ G(E)$ reflects the expectation that the effective gravitational
coupling will
depend on the energy scale and will satisfy a renormalization group
equation.
Similarly we expect an energy dependent cosmological constant $\Lambda
(E)$.

As in the usual theory, these equations must satisfy
a number of consistency conditions: Bianchi's identities.
This leads to local energy conservation and the geodesic equation.

\section{Modified $FRW$ solutions}

We illustrate the basic principles of this new theory with the
simplest cosmological solutions. We begin with flat $FRW$
solutions, whose family of metrics are given according to
(\ref{metric}) and the usual symmetry requirements, by,
\f\label{friedmet} ds^2 (E) = -{ dt^2\over f^{2}(E)} + { a^2 (t)
\over g^2 (E) } \gamma _{ij}dx^i dx^j \ff where $\gamma _{ij}$
represents the spatially homogeneous and isotropic metric of a
sphere (positive curvature $K=1$), pseudo-sphere (with negative
curvature $K=-1$), or euclidean space ($K=0$, so that $\gamma
_{ij}=\delta_{ij}$).

The only non-vanishing components of the associated connection
are:
\bea
\Gamma^0_{ij}&=&{\left( f\over g\right)}^2 a\dot a \gamma_{ij}\\
\Gamma^i_{0j}&=&\delta^i_j {\dot a \over a}\\
\Gamma^i_{jk}&=&{\tilde \Gamma}^i_{jk} \eea where ${\tilde
\Gamma}^i_{jk}$ are the standard spatial connection coefficients.
This leads to Riemann tensor components: \bea
R^0_{\; i0j}&=&{\left( f\over g\right)}^2 a\ddot a \gamma_{ij}\\
R^i_{\; 00j}&=&\delta^i_j {\dot a \over a}\\
R^i_{\; jkm}&=&{\tilde R^i_{\; jkm}}+{\left( f\over
g\right)}^2\dot a^2 (\delta^i_k\gamma_{jm}-\delta^i_m\gamma_{jk})
\eea with all other components zero or trivially derived from
these ones using the symmetries of the Riemann tensor. The
non-trivial Ricci tensor components are: \bea
R_{00}&=&-3{\ddot a\over a}\\
R_{ij}&=&\gamma_{ij}{\left( {\left( f\over g\right)}^2
(\ddot a a + 2 \dot a^2) + 2K\right)}
\eea

The energy momentum tensor has a perfect fluid form, \f T_{\mu
\nu} = \rho u_\mu u_\nu + p (g_{\mu \nu } + u_\mu u_\nu ) \ff but
now  $u_\mu$ depends on $E$ through $u_\mu = (f^{-1}(E), 0,0,0)$,
since it is a unit vector, $ u_\mu u_\nu g^{\mu \nu} =-1$, and $g$
depends on $E$. The energy density and pressure, $\rho$ and $p$,
do not depend on $E$, but they do depend on the temperature, and
through it on $a$. However the equation of state, $p=w\rho$, is
modified due to the deformed energy momentum relations, as
discussed extensively in \cite{ncvsl,ncinfl}.

The Einstein's equations become,
\f\label{f1}
\left ( {\dot{a} \over a} \right )^2 = {8\pi\over 3}
 G(E) {\rho\over f^2} -{K\over a^2}{\left( g\over f\right)}^2 + {\Lambda
(E)\over 3} \ff \f\label{f2} {\ddot{a} \over a} = -{4\pi\over 3}
G(E) {\rho+3p \over f^2} + {\Lambda (E)\over 3} \ff These may be
combined to produce a conservation equation: \be \dot\rho+3{\dot
a\over a}(\rho+p)=0 \ee

It is simple to investigate the consequences in some simple cases.
For example, assume that $G$ and $\Lambda$ have in fact the same
energy dependence, \f G(E) = h^2(E) G, \ \ \  \Lambda (E) = h^2(E)
\Lambda. \ff The effect of the energy dependence can be mocked up
by introducing an energy dependent time coordinate \f \tau (E) =
{h(E) \over f(E)} t \ff Then the effect is that a high energy
quanta of energy $E$, high enough to make the functions $h$ and
$f$ differ from unity,
 sees itself to be in a universe which is older or younger than a
 low energy quanta. In this way the horizon problem may be solved,
 or worsened, depending on the forms of the functions.

Indeed this theory of gravity opens up the doors for a new type of
solution to the horizon problem. Recall that the comoving horizon
is given by \be \label{comhor}r_h={cH^{-1}\over a} \ee and that
the horizon problem is that this quantity increases in time. Two
possible solutions, rendering $r_h$ a decreasing function of time,
are accelerated expansion ($\ddot a>0$) and a decreasing speed of
light ($\dot c<0$). Our theory may realize either of these
solutions~\cite{ncinfl,ncvsl}. As shown in \cite{ncinfl},
radiation subject to deformed dispersion relations may satisfy
$\rho+3p<0$, leading to inflation. Also, since the speed of light
may be obtained by setting $ds^2=0$ in (\ref{ds2}) we find \be
c={dx\over dt}={g\over f} \ee (note that we are setting the low
energy value of $c_0=1$; also $dx^0=c_0 dt$). If $f\neq g$, we may
then realize the VSL scenario~\cite{ncvsl}.

But more interestingly, there is a third solution. It is possible
to solve the horizon problem with decelerated expansion and
constant speed of light simply  because the metric is energy
dependent. Consider the various factors in (\ref{comhor}). The
Hubble parameter, $H=\dot a/a$, is energy independent because a
factor of $g$ appears in both numerator and denominator. The speed
of light $c(E)=g/f$, and $a$ is replaced by $a/g$ (see
(\ref{friedmet})). Hence the comoving horizon is \be
r_h={c(E)H^{-1}\over a/g} \ee If $g(E)\rightarrow 0$ fast enough
at early times  the comoving region containing the whole
observable universe nowadays may therefore be much smaller than
expected, solving the horizon problem. The conversion factor
between comoving and proper distances in an expanding universe is
energy dependent, and this is enough to bring the whole observed
universe together at high energies.

 An interesting case is the dispersion relation proposed in
\cite{leejoao}: \be f=g={1\over 1+\lambda E} \ee It does not
produce a varying $c$, and yet it may solve the horizon problem.

We shall return to this cosmological scenario in a future
publication. Our theory of gravity justifies adhoc assumptions
used in \cite{ncvsl,pogo}. Thus we may convert our theory into a
scenario for structure formation based on thermal
fluctuations~\cite{pogo}.

\section{The modified Schwarzschild solution}

We begin with the general spherically symmetric metric. Given that
the metric is energy dependent, when we specialize the coordinates to
spherically symmetric form, we may write the metric either in energy
dependent coordinates, or energy independent coordinates. The energy
independent coordinates will be denoted $\tilde{r}, \tilde{t},\theta,
\phi$. The time and radial coordinates can also absorb energy
dependence, leading to energy
dependent coordinates, which will be denoted $r(E), t(E)$. The angular
coordinates $\theta, ,\phi$ are always energy independent.
The energy independent coordiates will coincide with the energy
dependent coordinates in the limit $E/E_{Pl} \rightarrow 0$.

In terms of energy independent
coordinates the most general form for a spherically symmetric metric
is
\f
ds^2 = - { \tilde{F}(\tilde{r}) \over f^2 (E) } d\tilde{t}^2 +
{\tilde{H}(\tilde{r}) \over g^2 (E) } d\tilde{r}^2 + {\tilde{r}^2
\over g^2(E)} d\Omega^2
\label{first}
\ff

$\tilde{r}$ is hence the area coordinate, defined so it is
proportional to the square root of a physical area measured at that
radius, as
measured by observers in the
limit $E/E_{Pl} \rightarrow 0$.  Now, because the metric is energy
dependent, an observer using quanta of energy $E$ will measure a
sphere at constant $\tilde{r}$ to have a different area than an
observer using probes of energg $E=0$. We see that the area coordinate
appropriate to measurements made by quanta of energy $E$ is
\f
r(E) = {\tilde{r} \over g(E)}
\ff
We can then define new energy dependent functions,
\f
F(r(E),E)= \tilde{F}(\tilde{r}), \ \ \  H(r(E),E)= \tilde{H}(\tilde{r})
\label{rescale}
\ff
It is also natural to introduce an energy dependent time coordinate
\f
t(E)= {\tilde{t} \over f(E)}
\ff
The metric then takes the form,
\f
ds^2 = - {F}(r(E),E)  dt(E)^2 +
H(r(E),E)   dr(E)^2 + r(E)^2  d\Omega^2
\ff
This is then a one parameter family of metrics, each in the standard
form. By Birkoff's theorem, we must have, for each $E$
\f
F(r(E),E)= H^{-1}(r(E),E) = \left ( 1- {C(E) \over r(E)}
\right )
\ff
where the constant of integration $C(E)$ is now energy dependent.

However, we determine the energy dependence of the constant of
integration, because we recall that by (\ref{rescale}) all the
energy dependence in the energy-independent coordinates must be in
the functions $f$ and $g$ in the original form (\ref{first}). Hece
we have, \f F(r(E),E)=\left ( 1- {C(E)g(E)  \over \tilde{r} }
\right ) = \tilde{F}(\tilde{r}) \ff Thus, using the fact that at
$E=0$ we must have $C=2G(0)M$, \f C(E)= {2G(0)M \over g(E)} \ff
where $M$ denotes the energy independent mass. Thus, in energy
independent coordinates the modified Schwarzschild metric must take
the form, \f ds^2= -{ \left ( 1- { 2G(0)M \over \tilde{r}} \right
) \over f^2(E)} d\tilde{t}^2  + { 1\over g^2(E) \left ( 1- {
2G(0)M \over \tilde{r}} \right )    } d\tilde{r}^2  + {\tilde{r}^2
\over g^2 (E)} d\Omega^2 \ff In particular we see that the
position of the horizon, in fixed, energy independent coordinates,
is fixed at the usual place. However, the area of the horizon is
then energy dependent. The implications of this result for black
hole thermodynamics, and photon dynamics around the horizon, 
are currently being investigated.

\section{The Newtonian limit}

To check the Newtonian and linear limits we expand around the metric
of modified or doubly special relativity to find,
\f
g_{\mu \nu}= \eta (E)_{\mu \nu} + h(E)_{\mu \nu}
\ff
where the modified Minkowski metric is
$\eta (E)_{\mu \nu} = diag ( -f^{-2}, g^{-2},g^{-2},g^{-2})$.
We define the usual trace reversed coordiate
\f
\hbar_{\mu \nu}= h_{\mu \nu} - {1\over 2} \eta_{\mu \nu} h
\ff
where $h=\eta^{\mu \nu} h_{\mu \nu}$ is easily seen to be energy
independent. Fixing the standard gauge conditions $\eta^{\mu \nu}
\partial_\mu h_{\nu \lambda}=0$ we find that the linearized Einstein
eqations reduce to
\f
\eta^{\mu \nu}(E) \partial_\mu \partial_\nu h_{00} = - 16 \pi G(E)
T_{00}
\ff
Let us write,
\f
h_{00} = {\Phi \over f^2(E)}
\ff
where $\Phi$ is the usual Newtonian gravitational potential. We note
that this form is required by the correspondence principle.

Using again the modified perfect fluid form of the energy momentum
tensor we find that, for static fields, \f \delta^{ij}
\partial_i\partial_j \Phi = - 16\pi G(E) g(E)^2 \rho \ff The
energy dependence $g(E)$ corresponds to the fact that the
euclidean coordiates in which the spatial metric takes the form
$g_{ij}= diag(1,1,1)$ are energy dependent. Alternatively, the
coordinate differentials dual to $p_i = g(E) \partial_i$ are
$e^i(E) =dx^i /g(E)$. If we transform to energy independent
coordinates \f \tilde{x}^i = x^i g(E) \ff Then we find in these
coordinates, \f \delta^{ij} \partial_i\partial_j \Phi = - 16\pi
G(E) \rho \ff However, the Newtonian limit corresponds to the
limit $c\rightarrow \infty$. In this same limit $E_{Pl} =
\sqrt{\hbar c^5 /G} \rightarrow \infty$. Hence, by the
correspondence principle $G(E/E_{Pl}) \rightarrow G(0)=G$.  Hence
we do find, in energy independent coordinates, the correct
Newtonian limit, \f \delta^{ij} \partial_i\partial_j \Phi = -
16\pi G \rho \ff proving the weak field consistency of the theory.

\section{Conclusions}

We have proposed a way of incorporating spacetime curvature -- and thus
gravity -- into non-linear realizations of relativity (also known
as DSR). The gravity theory proposed in this paper is closely
related to one view of how to dualize doubly special relativity.
For a number of historical reasons (the theory is motivated by
cosmic ray kinematics) non-linear relativity was first defined in
momentum space. With loss of linearity duals no longer mimic each
other, so that recovering position space in these theories is
highly non-trivial.

One possible strategy is to impose linearity to the contraction
between position and momentum space. This is equivalent to
requiring that field theories still have plane wave
solutions~\cite{djj}. This requirement is strong enough to fully
fix the transformation laws of position space, which must still be
linear, but acquire an energy dependence. Consequently a flat
quadratic metric may still be defined, but the energy dependence
of the new Lorentz transformations propagates into the metric. It
is this ``rainbow'' metric that we gauged in this paper, using the
usual techniques of differential geometry.

We explored some solutions to this theory of gravity, namely the
cosmological, black hole, and weak field solutions. We found that
cosmological distances, in an expanding universe, become energy
dependent. Thus the physical distance associated with a given
comoving distance depends on the energy scale at which it is
measured. Seen in another way, the age of the universe is energy
dependent. We used this fact to show how the  horizon problem may
be solved without inflation or a varying speed of light. We also
considered the counterpart of the Schwarzschild solution, and found
that the area of the horizon is energy dependent. This result may
have  important implications for black hole thermodynamics.

Other theories of gravity would follow from different realizations
of position space. If the Lorentz group is non-linearly realized
{\it in position space}~\cite{djj}, gravity follows 
from gauging a symmetry which is non-linearly realized. If
position space is non-commutative, then yet another theory 
of gravity would emerge (however see \cite{eli} for a possible
obstruction). 
These alternative paths should be pursued
and compared with the theory proposed in this paper.

\section*{Acknowledgements}We would like to thank G. Amelino-Camelia,
J-L. Lehners, F. Markopoulou and J. Kowalski-Gilkman for useful discussions. 
JM thanks the Perimeter Institute for hospitality and LS
the Jesse Phillips Foundation for generous support which made this work
possible.

\end{document}